\documentstyle[amsfonts,12pt]{article}
\input tcilatex
\begin{document}

\title{Are gluons massive ?}
\author{Amjad Hussain Shah Gilani \\
National Center for Physics\\
Quaid-i-Azam University\\
Islamabad 45320 Pakistan}
\date{April 2004}
\maketitle

\begin{abstract}
It is claimed that only one gluon is massless and the other seven gluons are
massive. Out of eight gluons, six are colored and two are neutral. Among
neutral gluons, one is massless and other one is massive. Massive neutral
gluon is heavier than the colored gluons. Gluons can only be predicted by
set theory but not by $SU\left( 3\right) $.

[Report No.: NCP-QAU/2004-004, hep-ph/0404026]
\end{abstract}

Recently a strong indication for a deviation from the standard model (SM)
has been obtained by PIBETA Collaboration \cite{pibeta} in the $\pi
^{+}\rightarrow e^{+}\nu \gamma $ decay. A brief discussion on this
observation is given in Ref. \cite{refbrief}. A variety of literature is
available which give the indication of New Physics \cite{0312025}. This
motivates the possibility that there may be some physical phenomenon which
did not describe correctly and needs revision.

The SM is an incomplete theory: some kind of new physics is required in
order to understand the patterns of quark and lepton masses and mixings, and
generally to understand flavor dynamics \cite{CERN-2003-002-corr}. There are
strong theoretical arguments suggesting that new physics cannot be far from
the electroweak scale. Therefore, the determination of
Cabbibo-Kobayashi-Maskawa (CKM) matrix \cite{CKM1,CKM2} that parametrizes
the weak charged current interactions of quarks is currently a central theme
in particle physics. Test of its structure, conveniently represented by the
unitarity triangle, have to be performed; they will allow a precision
determination of the SM contribuions to various observables and possibly
reveal the onset of new physics contributions. Indeed, the four parameters
of this matrix govern all flavour changing transitions involving quarks in
the SM. These include tree level decays mediated by $W$ bosons, which are
essentially unaffected by new physics contributions. The flavour changing
neutral current (FCNC) transitions responsible for rare and CP violating
decays in the SM, which involve gluons, are sensitive probes of new physics.

The Standard Model (SM), all matter is made up of three kinds of elementary
particles: leptons, quarks, and gauge bosons (mediators). There are six
leptons with three generations. Similarly, there are six quarks, each quark
and antiquark comes in three color charges (red, green, blue), so there are
36 of them \cite{griffiths}. Before the Glashow-Salam-Weinberg (GSW) theory,
there were four fundamental forces in nature: strong, weak, electromagnetic,
and gravitational. The GSW model treats weak and electromagnetic interations
as different manifestation of a single electroweak force, and in this sense
the four forces reduce to three. Each of these forces is mediated by the
exchange of particle. The gravitational force is mediated by Graviton,
electroweak force by photon and intermediate vector bosons, and strong force
by gluons; and the Casimir force\footnote{%
The Casimir force also exist in nature which is the manifestation of
zero-point energy \cite{casimir}.}.

$
\begin{tabular}{|l|l|l|}
\hline
Force & Charge & Mediators (gauge bosons) \\ \hline
Casimir & \{ \thinspace \} & ? \\ \hline
Gravitational & \{0\} & Graviton \\ \hline
Electroweak & \{+, $-$\} & \{Photon, $W^{+}$, $W^{-}$,$Z^0$\} \\ \hline
Strong & red, green, blue & eight colored and massless gluons \cite
{griffiths} \\ \hline
\end{tabular}
$

If we look upon the above table, the mediators of their respective forces
has a certain relation which we can describe by set theory as discussed in
appendix \ref{appa}.

\noindent $%
\begin{tabular}{|l|l|l|l|}
\hline
Charges & No. of mediators & Set of charges & Subsets of mediators \\ \hline
0 & $2^0=1$ & \{ \} & Casimirion \\ \hline
1 & $2^1=2$ & $\left\{ 0\right\} $ & ${\frak g}$, ${\cal G}$ \\ \hline
2 & $2^2=4$ & $\left\{ +,-\right\} $ & $\gamma ,W^{+},W^{-},Z^0$ \\ \hline
3 & $2^3=8$ & $\left\{ r,g,b\right\} $ & $%
g_0,g_r,g_g,g_b,g_{rg},g_{rb},g_{gb},g_{rgb}$ \\ \hline
\end{tabular}
$

\noindent where $r,g,b$ stands for red, green, blue respectively i.e. the
color charges. We restrict here to the discussion of strong force while the
discussion on Casimir and gravitational forces are out of scope of this
article\footnote{%
As Casimir force is the quantum fluctuation of the vacuum and hence, no
charge exist there. We give an empty set to its charge. So, only one
massless mediator is required for such interactions. We suggest its name as
Casimirion. Now, we come to the gravitational force. All the matter has
charge neutral, we cannot associate an empty set to the charge upon the
matter. Therefore, we assign a `0' charge upon the matter. This set has two
subsets and both are improper subsets (see appendix \ref{appa}). This means
that there must be two mediators to describe gravitational interactions. As
nature never leave the gaps unfilled. So, we predict that there must be two
neutral gravitons, one massless ${\frak {g}}$ and other massive ${\cal {G}}$.%
}. The mediators $g_0,g_r,g_g,g_b,g_{rg},g_{rb},g_{gb},g_{rgb}$ are the
gluons for the strong force. We claim that only one gluon i.e. $g_0$ which
associates to empty subset is massless as photon $\left( \gamma \right) $ is
associated to empty subset in the case of electroweak mediators. The
remaining gluons $g_r,g_g,g_b,g_{rg},g_{rb},g_{gb},g_{rgb}$ are massive as
the intermediate vector bosons $W^{+},W^{-},Z^0$. The gluons $g_0$ and $%
g_{rgb}\left( \equiv G_0\right) $ are neutral. The gluons $%
g_{rg},g_{rb},g_{gb}$ are respectively equivalent to $g_{\bar{b}},g_{\bar{g}%
},g_{\bar{r}}$ as 
\begin{equation}
\begin{tabular}{llllll}
$\bar{r}$ & $\equiv $ & $gb,$ & $r$ & $\equiv $ & $\bar{g}\bar{b},$ \\ 
$\bar{g}$ & $\equiv $ & $rb,$ & $g$ & $\equiv $ & $\bar{r}\bar{b},$ \\ 
$\bar{b}$ & $\equiv $ & $rg,$ & $b$ & $\equiv $ & $\bar{r}\bar{g}.$%
\end{tabular}
\label{cantic}
\end{equation}

Let us give an example to support our argument, on page 280 of Ref. \cite
{griffiths}, a red quark turned into a blue quark, emitting a red-antiblue
gluon. Let us concentrate on the charge of the gluon in the above example.
The charge antiblue, $\bar{b}=rg$, is composed of `red and green', while the
charge over the gluon is `red-antiblue' equivalent to `red-red-green' which
does not obey the group property because `red' is repeated twicely, as we
define the anticolors in Eq. (\ref{cantic}). This argues that a gluon will
never carry a charge like $r\bar{b}$ etc. It will only carry a color or
anti-color. We claim that in the above example a red quark will turn into
blue quark, emitting a green gluon. We also point out that 
\begin{eqnarray*}
m_{g_r} &=&m_{g_{gb}}\left( =m_{g_{\bar{r}}}\right) \\
m_{g_g} &=&m_{g_{rb}}\left( =m_{g_{\bar{g}}}\right) \\
m_{g_b} &=&m_{g_{rg}}\left( =m_{g_{\bar{b}}}\right)
\end{eqnarray*}
The gluon $G_0$ will be massive than the gluons $g_r,g_g,g_b,g_{\bar{b}},g_{%
\bar{g}},g_{\bar{r}}$ as the neutral vector boson $Z^0$ is massive to
charged vector boson $W^{\pm }$. That is 
\[
m_{G_0}>m_{g_i},\,\,\,i=r,g,b,\bar{b},\bar{g},\bar{r} 
\]

We conclude that one gluon is massless and the seven gluon are massive. Of
them massless gluon is neutral and one massive gluon is color singlet. The
rest of the six gluons are colored and massive, their mass relations are
also given.

The massive gluons can interact with each other in similar way as the vector
bosons interact, while the massless gluon play the similar role as photon.
We can divide QCD in two branches, one which deals the interaction mediated
by massless gluon and the other which deals with the interactions by the
rest of gluons, say chromo-magnetic and chromo-weak, respectively. The names
suggested on the basis of electro-magnetic and weak theory. The marriage of
chromo-magnetic and chromo-weak theories will result in QCD. The
chromo-magnetic and chromo-weak interactions will collectively be called as
strong interactons. We summarize our findings as:

\begin{itemize}
\item  color-induced interactions between quarks are mediated by gluons and
electroweakly neutral spin-1 particles,

\item  only one gluon is massless and remaining seven gluons are massive,

\item  six gluons are colored, three carry color charge and three carry
anticolor charge,

\item  two gluons are color neutral, one massless and one massive,

\item  neutral massive gluon is heavier than the colored gluons,

\item  the colored quarks emit and absorb massless gluon in the same way as
the electrically charged particles emit and absorb photons,

\item  the colored quarks emit and absorb massive colored and neutral gluons
in the same way as the electrically charged particles emit and absorb vector
bosons $W^{\pm }$ and $Z^0$ respectively,

\item  the gluons can be predicted by set theory but not by $SU\left(
3\right) $ in analogy to electroweak mediators $\gamma ,W^{+},W^{-},Z^0$.
\end{itemize}

\appendix

\section{Sets and their subsets \label{appa}}

In the late ninteenth century Georg Cantor \cite{cantor} was the first to
realize the potential usefulness of investigating properties of sets in
general as distinct from properties of the elements that comprise them \cite
{SSEpp}. All objects can be defined in terms of sets.

The words set and element are undefined terms of set theory just as
sentence, true and false are undefined terms of logic. The founder of set
theory, George Cantor, suggested imagining a set as a ``collection of
definite and separate objects of our intution or thought. These objects are
called elements''. A set is completely determined by its elements; the order
in which the elements are listed is irrelevant. 
\[
\begin{tabular}{llll}
Sets & No. of & Subsets &  \\ 
& subsets &  &  \\ 
$\left\{ \,\,\right\} $ & $2^0=1$ & $\left\{ \,\,\right\} $ &  \\ 
$\left\{ 0\right\} $ & $2^1=2$ & $\left\{ \,\,\right\} ,\left\{ 0\right\} $
&  \\ 
$\left\{ +,-\right\} $ & $2^2=4$ & $\left\{ \,\,\right\} ,\left\{ +\right\}
,\left\{ -\right\} ,\left\{ +,-\right\} $ &  \\ 
$\left\{ r,g,b\right\} $ & $2^3=8$ & $\left\{ \,\,\right\} ,\left\{
r\right\} ,\left\{ g\right\} ,\left\{ b\right\} ,$ &  \\ 
&  & $\left\{ r,g\right\} ,\left\{ r,b\right\} ,\left\{ g,b\right\} ,\left\{
r,g,b\right\} $ & 
\end{tabular}
\]

A basic relation between sets is that of subset. There are two types of
subsets i.e. proper and improper subset. The empty set and set itself are
improper subsets of a set. To check whether one finite set is a subset of a
given set. If any element of a set is not found to equal any element of the
given set. Then, the set is not a subset of the given set.

Hofstadter points out that when you start a mathematical argument with if,
let, or suppose, you are stepping in a fantasy world where not only are all
the facts of the real world true but whatever you are supposing is also true 
\cite{Godel}. Once you are in that world, you can suppose something else.
That sends you in subfantasy world where not only is everything in the
fantasy world true but also the newthings you are supposing. Of course you
can continue stepping into new subfantasy worlds in this way indefinitely.
You return one level closer to the real world each time you derive a
conclusion that makes a whole if -- then or universal statement true. Your
aim in a proof is to continue deriving such conclusions until you return to
the world from which you made your first supposition. So, in Hofstadter's
terms, the author invites the reader to enter in fantasy world where one
statement is known to be true and the other to provein this fantasy world.

Thanks to the warm discussion between Professor Guido Altarelli \cite
{altarelli} and Professor Pervez Hoodbhoy \cite{hoodbhoy} during the 3rd
Workshop on Particle Physics held in Islamabad, Pakistan which motivate us
to write this article. Thanks to Professor Ahmed Ali for suggesting me a
problem ``$B\rightarrow K^{**}\gamma $ decay'' to solve in QCD factorization
approach following the pattern of his paper \cite{0105302} and Ref. \cite
{safir} during his visit to Pakistan, March 2002. While reviewing his paper 
\cite{0105302}, a mistake was found in reproducing Figures 7 and 8 of Ref. 
\cite{0105302}, which took me more than three months to exactly trace out
where the mistake was and the authors \cite{0105302} are agreed with my
observation. During the above period in which I was reviewing the paper \cite
{0105302}, I repeated the color concept calculations includes color factor,
color upon the gluons, etc., which gave me another confusion but was unable
to sketch it. Third workshop on particle physics suddenly solve the mystry
which results this article.

Acknowledgements: This work is supported by Pakistan Council for Science and
Technology. We would like to thank the invited lecturers of 3rd Workshop on
Particle Physics for their stimulating lectures \cite{altarelli,3wpp}. G
thanks to Professor Ahmed Ali, Professor Guido Altarelli and Professor John
Ellis for discussions during their visits to Pakistan. Thanks are also due
to Shoaib, Jamil, Irfan, Ijaz, Mariam, Almas, Shahid, Shabbar, Naeem, Usman,
Rizwana and Hammad Chudhary for discussions. Thanks to Dr. Asif Ali for
providing me reference \cite{SSEpp}. I cannot find proper words to pay my
thanks to my teachers Professor Riazuddin and Professor Fayyazuddin for
their encourgment and help in building concepts through elementary level.

\end{document}